\documentstyle[12pt]{article}
\begin{document}

\setlength{\topmargin}{-.5in}
\setlength{\textheight}{9in}
\setlength{\parskip}{.3cm}
\setlength{\belowdisplayskip}{.3cm}
\setlength{\abovedisplayskip}{-.3cm}
\setlength{\belowdisplayshortskip}{.3cm}
\setlength{\abovedisplayshortskip}{-.3cm}
\setlength{\jot}{-.1cm}
\setlength{\baselineskip}{13pt}
\newcommand{\ii}{\'{\i}}
\newcommand{\be}{\begin{equation}}
\newcommand{\ee}{\end{equation}}
\newcommand{\bea}{\begin{eqnarray}}
\newcommand{\eea}{\end{eqnarray}}
\newcommand{\ups}{\upsilon}
\newcommand{\ra} {\rightarrow}
\newcommand{\vp} {\varphi}


\hrule 
\vskip 5.cm

\centerline{{\Large Topologically Massive Models from Higgs 
Mechanism}}
\vskip.5cm
\centerline{Talk given at " I Congreso Venezolano de F\'{\i}sica,  
ULA M\'erida, 1997"}
\vskip.5cm 
\centerline{{\large A. De Castro and A. Restuccia}}
\vskip .5cm
\centerline{\it Universidad Sim\'{o}n Bol\'{\i}var, Departamento de 
F\'{\i}sica.}

\centerline{\it Apartado postal 89000, Caracas 1080-A, Venezuela.}
\centerline{\it e-mail: alex@fis.usb.ve, arestu@usb.ve}

\vskip 1cm
{\bf Abstract}
\begin{quotation}
\noindent{\small A  Higgs mechanism for Abelian theories over 
non-trivial background flat connections is proposed.  It is found that the mass 
generated for the spin $1$ excitation is the same as the one obtained 
from the standard Higgs mechanism over trivial backgrounds, however, 
the dynamical structure of the action for the Higgs scalar is 
completely different from the usual approach. There is a topological 
contribution to the mass term of the Higgs field. After functional 
integration over all backgrounds, it is shown that the action for the 
massive spin $1$ excitation is dual to the Topologically Massive 
Models in any dimension.}
\end{quotation}

\newpage
\noindent{\Large{\bf{Introduction}}}
\vskip .5cm

The Higgs Mechanism of symmetry breaking, has been used very 
successfully as a mechanism of generating mass to gauge fields. For 
small coupling the quantum Hilbert space of a field theory is 
constructed by expanding around the classical minima of the potential. 
In the standard application of the Higgs mechanism to a massless 
gauge field coupled to scalar field the absolute minima of the 
potential correspond to $\left<\Phi\right>=\ups e^{i\alpha}$,   
($\ups$  constant) and to a  potential ${\bf A}$ gauge equivalent to 
zero. In other cases mainly for $SU(2)$ gauge potentials, because of  
terms of the type $Tr\left(F_{ij}F_{ij}\right)$ and non-trivial 
structure of the interacting terms for the scalar fields in the 
Hamiltonian, classical minima correspond to flat connections together 
with scalar fields satisfying some restrictions.  For example in the  
Witten's topological field theory{\cite{re:(1) witten}} describing 
Donaldson's invariants the scalar fields satisfy 
$D\Phi=D\Psi=\left[\Phi,\Psi\right]=0$.  Similar terms for the scalar 
fields occur for the $N=2$ SUSY Yang-Mills theory. The moduli space 
of minima is then very rich  and allows interesting relations between 
topological aspects of the base manifold and physical observable of 
the resulting field theory. After the successful use of the Higgs 
mechanism to the construction of the Standard Model, several 
interesting theories have been formulated which describe spin 1 massive 
excitations in terms of a gauge invariant field theory. In the 
Topological Massive{\cite{re:(2) Deser}} theory in 3 dimensions the 
``topological  mass'' is introduced via a Chern-Simon term in the 
Lagrangian, several extensions of this theory to four and higher 
dimensions have been presented in the literature in terms of 
antisymmetric fields and $BF$ interacting terms. Also much work has 
been developed to understand the origin of new mechanism of generating mass to 
gauge field theories {\cite{re:(3) Cremmer-Scherk}}. In {\cite{re:(4) 
Mackenzie}} it was shown that the quantum contribution from one loop 
diagrams to the effective action in theories with four fermions 
coupling in four dimensions exactly reproduces the topological terms 
of an antisymmetric field action describing the new topological 
photon mass generation. They appear in a similar way as the 
Chern-Simon terms are generated by parity violating massive fermions 
in $2+1$ dimensions{\cite{re:(5) Redlich}}. All these theories 
describe massive excitations over trivial fiber bundles.

  The new relevant points in the topological mass generating approach are the 
gauge invariant formulation together with the existence of new 
topological sectors in these theories {\cite{re:(6) 
jorge-pio-alvaro}}{\cite{re:(7) Pio-Leal-alvaro}}, which may be 
understood in terms of topological theories. In most of the cases in 
terms of $BF$ topological theories. The main ingredient in the 
formalism is the presence of non-trivial flat connections (Bohm 
Aharonov potentials) which carries into the theory the non-trivial 
topological aspects of the base manifold. Because of the presence of 
these non-trivial flat connections and the remark that in the 
standard Higgs mechanism the minima of the Hamiltonian impose the 
condition of trivial connections, it is thought that these massive 
gauge theories are related to a different mass generating mechanism.

Since this new mechanism of photon mass generation is topological in 
nature we will look for a modification of the variational problem of 
the standard Lagrangian  of massless gauge fields coupled to scalars 
which may take into account topological aspect of the base manifold. 
 This modification allows the introduction of non-trivial flat 
connections in the description of the background.  

We start from the standard Lagrangian for a massless gauge field 
coupled to scalars, formulated on a topological non-trivial base 
manifold. We then show the well known result that the absolute minima 
of the action under arbitrary variations imposes the condition to the 
connection to be pure gauge. We then introduce a restricted space of 
variations and consider stationary points on this new space. We show 
that there are minima associated to non-trivial flat connections. We 
then follow the Higgs mechanism over this background and obtain the 
field equations for the Higgs scalar and for the massive gauge field. 
The mass of the spin $1$ excitation is the same as in the standard 
Higgs approach  while the mass of the Higgs scalar has a 
``topological" contribution from the nontrivial flat connections on 
the background. Finally  we perform a duality transformation to obtain 
the TM theory in 3 dim and its generalizations to higher dimensions 
involving antisymmetric fields and $BF$ interacting terms. 

\vskip .5cm
\noindent{\Large{\bf{Higgs Mechanism over Non-Trivial Flat\\
\vskip -.5cm
\noindent{}Connections}}}
\vskip .5cm

We start with the action of a complex scalar field interacting with 
an Abelian gauge field,

\be
S=\left<D_\mu\Phi\bar{D}^\mu\Phi^*-V(\Phi)+{\cal L}(A)\right>
\label{eq:accion de partida}
\ee

\noindent here $\left<{}\right>$ denotes integration on a $d$ 
dimensional space-time, $D_\mu=\partial_\mu+iA_\mu$, the potential
$V(\Phi )$ is given by

\be
V(\Phi)={\mu}^2({\Phi}^*\Phi)+\mid\lambda\mid({\Phi}^*\Phi)^2,
\ee

\noindent and ${\cal L}(A)$ stands for the Lagrangian associated to 
the $U(1)$ connection $A$. We will consider ${\cal L}(A)$ to be the 
Chern-Simons Lagrangian on $d=3$ or the Maxwell Lagrangian on 
$d\geq{}3$ dimensions. In any case the action is invariant up to 
total derivatives under the gauge transformation: $\Phi\ra 
e^{i\xi}\Phi$ and $A\ra A-d\xi$. 

We would like to extend the spontaneous symmetry breaking process in 
such a way as to include nontrivial flat connections  (one cannot 
gauge the background $A$ away). The extension is naturally suggested 
by the fact that the Hamiltonian shows an effective potential 
$V(\Phi{})_{eff}$ given by:

\be
V(\Phi)_{eff}=A_iA_i 
(\Phi^*\Phi)+\mu^2(\Phi^*\Phi)+\lambda(\Phi^*\Phi)^2
\ee

\noindent$i=1,2...,d-1$

Meaning that the gauge potential could contribute to the mass of the 
Higgs boson.
 
We shall begin our analysis showing that a naive study of a vacuum 
state with $A\neq{}0$ leads back to the usual Higgs solution $A=0$. 
After this, we will define the appropriate modifications that will end 
up in the solution for nontrivial topologies.

We denote the non-trivial vacuum expectation value  
$\left<\Phi\right>=\ups$, $\ups$ being a constant parameter, which 
without loosing generality we may take to be real.

The field equations for the action (\ref{eq:accion de partida}) are 
obtained by taking general variations with respect to $\Phi$, 
$\Phi^*$ and $A_\mu$. We are interested in solutions for the scalar 
and vector fields of the form:

\bea
\Phi_0&=&\ups e^{i\xi}\nonumber\\
\nonumber\\
A_\mu&=&\hat{A}_\mu\neq{}0 \hskip .5cm.
\label{eq:solphi}
\eea

\noindent{}where $\ups$ is a constant. 

Unrestricted variations of (\ref{eq:accion de partida}) with respect 
to $\Phi$ and $\Phi^*$ yield,

\bea
\partial_\mu\partial^\mu\Phi^*-i\partial_\mu{}A^\mu\Phi^*-2iA^\mu\partial_\mu\Phi^*-A_\mu{}A^\mu\Phi^*+\mu^2\Phi^*+2\lambda(\Phi\Phi^*)\Phi^*=0\nonumber\\
\nonumber\\
\partial_\mu\partial^\mu\Phi+i\partial_\mu{}A^\mu\Phi+2iA^\mu\partial_\mu\Phi-A_\mu{}A^\mu\Phi+\mu^2\Phi+2\lambda(\Phi\Phi^*)\Phi=0
\label{eq:eq de mov1}
\eea

After replacing (\ref{eq:solphi}) into (\ref{eq:eq de mov1}) we 
obtain for the background field $\hat{A}_\mu$.

\be
\partial_\mu\hat{A}^\mu=0
\label{eq:sol hatA1}
\ee

\be
-\hat{A}_\mu\hat{A}^\mu+\mu^2+2\lambda\ups^2=0 \hskip .5cm.
\label{eq:sol hatA2}
\ee

Unrestricted variations of (\ref{eq:accion de partida}) with  respect 
to $A_\mu$ yield,

\be
i\Phi\bar{D}^\mu\Phi^*-i\Phi^*D^\mu\Phi+\frac{\delta{\cal 
L}}{\delta{}A_\mu}=0 \hskip .5cm.
\label{eq:var con resp Amu}
\ee

After replacing (\ref{eq:solphi}) into (\ref{eq:var con resp Amu}) we 
get 

\be
2\ups^2\hat{A}^\mu+\left.\frac{\delta{\cal 
L}}{\delta{}A_\mu}\right|_{\hat{A}_\mu}=0
\label{eq:eval of hatA}
\ee

Since we are looking for the ground state of the theory, we are not 
only interested in a stationary point of the action but also in 
having a minimum of the Hamiltonian. If ${\cal L}(A)$ is the 
Lagrangian  of Maxwell theory this further  condition implies that 
$\hat{A}$ must be a flat connection, we will also consider $\hat{A}$ 
to be a flat connection in the case of Chern-Simons theory. Under 
such conditions, equation (\ref{eq:eval of hatA}) implies that:

\be
\hat{A}_\mu=0 \hskip .5cm.
\label{eq:pure gauge}
\ee

Which is the anticipated usual Higgs solution. In 
(\ref{eq:solphi})-(\ref{eq:pure gauge}) we used the Minkowski metric, 
however the results are valid for any (pseudo) Riemannian metric over 
the base manifold.

Our original interest lies in having a non-trivial background 
$\hat{A}_\mu$, from the above study, it is clear that we must modify 
the approach. We would like to have $\hat{A}_\mu$ a flat connection 
without further restrictions, this means that we must avoid the 
conditions imposed by equations (\ref{eq:sol hatA1}) and 
(\ref{eq:pure gauge}). Indeed, (\ref{eq:sol hatA1}) seems like a 
gauge fixing condition to $\hat{A}$ which is potentially conflicting 
with (\ref{eq:sol hatA2}).

In order to achieve a reasonable modification of the approach we 
consider stationary points of the action (\ref{eq:accion de partida}) 
with variations of the field subjected to certain restrictions, which 
for the scalar field we take

\be 
\frac{\delta\Phi}{\Phi_0}=\frac{\delta\Phi^*}{\Phi_0^*},
\label{eq:condition1}
\ee

and

\bea
d(\delta\Phi\hat{A})=0,\nonumber\\
\nonumber\\
\left<\delta\Phi\hat{A}^2\right>=0,
\label{eq:condition3}
\eea

\noindent{}where $d$ is the exterior derivative acting on the 1-form 
$\delta\Phi\hat{A}$, while for the vector field we take the global 
constraints

\bea
\left<\left[2\ups^2\delta{}A^\mu+\frac{\delta{\cal 
L}(\delta{}A)}{\delta{}A_\mu}\right]\hat{A}_\mu\right>&=&0,\nonumber\\     
\left<\delta\Phi\left[2\ups^2\delta{}A^\mu+\frac{\delta{\cal 
L}(\delta{}A)}{\delta{}A_\mu}\right]\hat{A}_\mu\right>&=&0\hskip.5cm.
\label{eq:condition2}  
\eea

This set of restrictions on the variations of the fields ensures that 
the stationary points $\Phi_0=\ups$ and $\hat{A}\neq{}0$ a flat 
connection are strict minima of the action (\ref{eq:accion de 
partida}). That is, these configurations are strict minima of the 
action for variations satisfying those constrains. It seems that such 
set is the unique one to achieve such goal. 

We thus obtain several minima depending on the existence of 
non-trivial flat connections on the base manifold. That is, depending 
on the non-trivial topology of the base manifold. The number of such 
minima is in one to one correspondence to the homomorphisms of the 
fundamental group of the base manifold into the structure group, 
$U(1)$ in our case.

All the results we are considering are valid over any (pseudo) 
Riemannian metric over the base manifold.

The perturbative expansion around the vacuum of the complex field 
$\Phi$ satisfying (\ref{eq:condition1})is:

\be
\Phi=e^{i\frac{\xi}{\ups}}(\ups+\eta)
\label{eq:parametrizacion de phi}
\ee

\noindent$\xi$ and $\eta$ being real fields, then we have

\be
D_\mu\Phi=e^{i\frac{\xi}{\ups}}\tilde{D}_\mu(\ups+\eta)\hskip.5cm.
\ee

Where

\bea 
\tilde{A}_\mu=A_\mu+\frac{1}{\ups}\partial_\mu\xi\nonumber\\
\nonumber\\
\tilde{D}_\mu=\partial_\mu+i\tilde{A}_\mu
\eea

\noindent{}and
 
\be
V(\Phi)={\mu}^2(\ups+\eta)^2+\lambda(\ups+\eta)^4\hskip.5cm.
\ee

Under a gauge transformation one has

\bea
\xi{}&\ra&\xi-\Lambda\nonumber\\
\nonumber\\
\tilde{A}&\ra& \tilde{A}+d\Lambda
\eea

\noindent{}so $\xi$ may be eliminated by a gauge transformation. 

We notice that (\ref{eq:parametrizacion de phi}) is the general 
solution to (\ref{eq:condition1}).

The action (\ref{eq:accion de partida}) then reduces to 

\be
S=\left<\partial_\mu\eta\partial^\mu\eta+\tilde{A}_\mu\tilde{A}^\mu(\ups+\eta)^2
-{\mu}^2(\ups+\eta)^2-\lambda(\ups+\eta)^4+{\cal L}(\tilde{A})\right>\hskip.5cm.
\label{eq:accion2}
\ee

Now we consider the decomposition of the $U(1)$ connection over a 
non-trivial flat background $\hat{A}_\mu$. 

We have 

\be
\tilde{A}_\mu=\hat{A}_\mu+a_\mu\hskip.5cm,
\label{eq: exp del A}
\ee

\noindent{}where $a_\mu$ satisfies (\ref{eq:condition2}) with 
$\delta{}A_\mu=a_\mu$.

\noindent(\ref{eq:accion2}) then reduces to

\bea
S&=&\left<\partial_\mu\eta\partial^\mu\eta+(\hat{A}^2-\mu^2)\eta^2-6\lambda\ups^2\eta^2+[2\ups(\hat{A}^2-\mu^2)-4\ups^3\lambda]\eta\right>\nonumber\\
\nonumber\\
&+&\left<4\ups\hat{A}_\mu a^\mu\eta+{\cal L}(a)+\ups^2 
a^2+2\ups^2\hat{A}_\mu a^\mu\right>
\label{eq:accion3}
\eea

\noindent{}plus terms independent of $\eta$ and $a$ and higher order 
terms which we have not written. 

The Higgs field $\eta$ and the massive vector field $a_\mu$ are 
coupled through the term $4\ups\hat{A}_\mu a^\mu\eta$ which depends 
on the background field $\hat{A_\mu}$. We notice also the presence of 
a non standart term $2\ups^2\hat{A}_\mu a _\mu$. If we consider 
$\eta$ and $a_\mu$ to be infinitesimal of the same order with 
respect to $\ups$ and $\hat{A}_\mu$ respectively in the expansions 
(\ref{eq:parametrizacion de phi}) and  (\ref{eq: exp del A}), we 
obtain from (\ref{eq:accion3}) two first order infinitesimal terms 

\be
2\ups[\hat{A}^2-\mu^2-2\ups^2\lambda]\eta+2\ups^2\hat{A}_\mu a^\mu,
\ee
which have to be annilated in order to have a consistent theory. 
Using (\ref{eq:condition2}), $2\ups^2\hat{A}_\mu a^\mu$ may be 
expressed as a total derivative. Using (\ref{eq:condition2}) and 
(\ref{eq:condition3}) the term $4\ups\hat{A}_\mu a^\mu\eta$ in the 
action (\ref{eq:accion3}) can also be written as a total derivative. 
The term $2\ups\hat{A}^2\eta$ is eliminated by the restriction 
(\ref{eq:condition3}). We take $\ups^2=\frac{-\mu^2}{2\lambda}$.

The constrained action is then equivalent to 

\bea
S&=&\left<\partial_\mu\eta\partial^\mu\eta+\left(\hat{A}^2-4\lambda\ups^2\right)\eta^2
\right>\nonumber\\
\nonumber\\
&+&\left<{\cal L}(a)+\ups^2a^2+(\theta + \rho\eta)\left(2\ups^2a_\mu+\frac{\delta{\cal 
L}}{\delta{}a^\mu}\right)\hat{A}^\mu\right>\hskip.5cm,
\eea

\noindent{}subject to (\ref{eq:condition3}). 

\noindent{}$\theta$ and $\rho$ are the Lagrange multipliers 
associated to (\ref{eq:condition2}). Since the constraints are global 
ones the Lagrange multipliers are constant.  (\ref{eq:condition1}) 
has already been implemented by using (\ref{eq:parametrizacion de 
phi})

The field equations for the massive vector field $a_\mu$ are 

\be
\frac{\delta{\cal L}}{\delta{}a_\mu}+2\ups^2a^\mu+2(\theta + 
\rho\eta)\ups^2\hat{A}^\mu+\frac{\delta^2{\cal 
L}}{\delta{}a_\mu\delta{}a_\nu}\left((\theta 
+\rho\eta)\hat{A}^\nu\right)=0
\label{eq:eq of mov for a1}
\ee

\bea
\left<\left(\frac{\delta{\cal 
L}}{\delta{}a_\mu}+2\ups^2a^\mu\right)\hat{A}_\mu\right>&=&0,\nonumber\\
\left<\eta\left(\frac{\delta{\cal 
L}}{\delta{}a_\mu}+2\ups^2a^\mu\right)\hat{A}_\mu\right>&=&0\hskip.5cm.
\label{eq:eq of mov for a2}
\eea

The last term in (\ref{eq:eq of mov for a1}) may be rewritten as 

\be 
{}^*d\left((\theta + \rho\eta)\hat{A}\right)
\ee

\noindent{}for tha case of 3 dim and the Chern-Simon Lagrangean, and

\be
{}^*d\,^*d\left((\theta + \rho\eta)\hat{A}\right)
\ee

\noindent{}for the case of the Maxwell Lagrangian in $D$ dimensions. 

In both cases, because of (\ref{eq:condition3}), the terms vanish. 
Using  (\ref{eq:eq of mov for a1}), (\ref{eq:eq of mov for a2}), 
(\ref{eq:condition3}) and (\ref{eq:condition2}) we get

\bea
\theta=0\nonumber\\
\\
\rho=0\nonumber
\eea

\noindent{}and the field equations reduce then to

\be
\frac{\delta{\cal L}}{\delta{}a_\mu}+2\ups^2a^\mu=0
\label{eq:ec de mov para vector}
\ee

\noindent{}where $a_\mu$ can be reexpressed in terms of the 
connection $A^\mu$ as

\be
a_\mu=A_\mu-\hat{A}_\mu\hskip.5cm.
\label{eq:a in terms of A}
\ee

(\ref{eq:ec de mov para vector}) shows that the generated mass for 
the spin $1$ excitation is exactly the same as in the standard Higgs 
mechanism over a trivial background. The scalar field $\eta$, 
however, is described by a completely different action 

\be
\left<\partial_\mu\eta\partial^\mu\eta+\left(\hat{A}^2-4\lambda\ups^2\right)\eta^2\right>
\ee

\noindent{}subject to (\ref{eq:condition3}).

The quadratic terms in this action show a distribution of mass 
depending on the topology of the base manifold since $\hat{A}$ is a 
closed one form.  Its contribution has the same sign  as  
$4\lambda\ups^2$ resulting in an increasement of the mass term. The 
constraint (\ref{eq:condition3}) severely restricts the space of 
solutions of the corresponding field equation.

\vskip .5cm
\noindent{\Large{\bf{Dual Formulation of the Spin $1$ Massive\\
\vskip -.5cm
\noindent{}Excitations}}}
\vskip .5cm

We would like now to analyze the action for the massive spin $1$ 
excitations. We will show that it is the dual formulation of the 
gauge covariant theories for spin $1$ excitations in term of 
antisymmetric fields, for the case in which ${\cal L}$ is the Maxwell 
Lagrangian in $d$ dimensions and of the Topological Massive theory in 
3 dimensions. From the point of view of the mass generating mechanism 
we started with a fixed background $\hat{A}$, we showed that for 
each  $\hat{A}$ we have a minimum of the restricted variational 
problem and we performed the perturbative analysis around it. We are 
now going to functionally integrate over all possible backgrounds. In 
this sense we will obtain an effective action for the massive spin 
$1$ excitation.

The action for the massive spin $1$ field is 

\be
\left<{\cal L}(a)+2\ups^2a_\mu{}a^\mu\right>
\label{eq:accion para a}
\ee

\noindent{}where $a_\mu$ is given by (\ref{eq:a in terms of A}),

\be 
a_\mu=A_\mu-\hat{A}_\mu
\ee

$A$ being the independent one form connection and $\hat{A}$ an 
independent closed one form over the base manifold. (\ref{eq:accion 
para a})may be rewritten as

\be
\left<{\cal 
L}(a)+2\ups^2\left(A-\hat{A}\right)\wedge{}^*\left(A-\hat{A}\right)\right>\hskip.5cm.
\ee

Where $*$ is the Hodge dual operation.

We will now follow the general construction of dual formulations 
described in {\cite{re:tesis de freddy}}{\cite{re:mario e isbelia}}. 
We introduce an independent 1-form $L$ globally defined over the base 
manifold satisfying

\be
dL=0
\label{eq:local condition}
\ee

\noindent{}and introduce this constraint into the action by using a 
Lagrange multiplier $d-2$ form $B$. We will assume that 
(\ref{eq:local condition}) is the only constraint on $L$ and then 
comment on the modification if a global restriction of the form

\be
\oint{}L=2\pi{}n
\label{eq:global condition}
\ee

\noindent{}is imposed on $L$. If there is no condition of type 
(\ref{eq:global condition}) then $B$ is a $d-2$ form globally defined 
over the base manifold. 

We then have

\be
\left<{\cal 
L}(A)+2\ups^2\left(A-L\right)\wedge\,^*\left(A-L\right)+iL\wedge{}dB\right>\hskip.5cm.
\label{eq:unconstrained act}
\ee

We may now integrate $L$ in the functional integral or equivalently 
take the field equation

\be
-4\ups^2\,^*\left(A-L\right)+idB=0
\ee

\noindent{}and eliminate $L$ in terms of $A$ and $B$. We obtain the 
quantum equivalent action

\be
\left<{\cal 
L}(A)+(-1)^d\frac{1}{8\ups^2}{}^*dB\wedge{}dB+iA\wedge{}dB\right>\hskip.5cm.
\label{eq:gauge invariant act}
\ee

If ${\cal L}(A)$ is the Maxwell Lagrangean over a 4-dimentional base 
manifold, (\ref{eq:gauge invariant act}) is exactly the gauge 
invariant action describing a massive spin $1$ excitation introduced 
in {\cite{re:(3) Cremmer-Scherk}. Its generalization to $d$ 
dimensions and equivalence to the Proca formulation was presented in 
the latest of {\cite{re:(3) Cremmer-Scherk} and also agrees with 
(\ref{eq:gauge invariant act}). If ${\cal L}(A)$ is the Chern-Simon 
lagrangean in $3$ dimensions and we functionally integrate on $A$ in 
(\ref{eq:gauge invariant act}) we end up with the Topological Massive 
action{\cite{re:(2) Deser}}, with the topological mass generated now 
by a Higgs mechanism on a non-trivial background.

If we consider a global condition of the form (\ref{eq:global 
condition}) then the correct lagrange multiplier $B$ is locally a 
$d-2$ form satisfying the global condition

\be
\oint_{\Sigma_{D-1}}dB=2\pi{}m\hskip.5cm.
\label{eq:charge cuantiz}
\ee

The constrains (\ref{eq:local condition}) and (\ref{eq:global 
condition}) are introduced into the unconstrained action 
(\ref{eq:unconstrained act}) with the same term

\be
iL\wedge{}dB
\ee

\noindent{}the general construction with global constrains was 
obtained in {\cite{re:tesis de freddy}}{\cite{re:mario e isbelia}}. 
The same local results are obtained as before, however the global 
structure of the field $B$ is different. Condition (\ref{eq:charge 
cuantiz}) works as a quantum  stabilizer for the different minima. 
The antisymmetric field $B$ satisfying (\ref{eq:charge cuantiz}) may 
have non-trivial transitions over the base manifold  associated to a 
higher order bundles as described in {\cite{re:mario e isbelia}}. If 
$B$ is a globally defined $d-2$ form then $m=0$ in (\ref{eq:gauge 
invariant act}).

\vskip .5cm
\noindent{\Large{\bf{Conclusions}}}
\vskip .5cm

We introduced a new variational principle which allow the 
implementation of the Higgs mechanism for Abelian theories over 
non-trivial background flat connections. The construction has 
similarities to the variational problem that allows, for integrable 
systems, to obtain the different multi-solitonic solutions as minima 
of constrained optimization problems constructed from conserved 
quantities.

The resulting actions for the massive spin $1$ excitations are gauged 
invariant and dual to the Topological Massive action in 3 dimensions 
and to its generalizations on higher dimensions in terms of 
antisymmetric fields. The approach shows how all these massive gauge 
actions may be obtained from massless gauge actions through the Higgs 
mechanism. The mass of the spin $1$ excitation is the same as the one 
obtained from the standard Higgs mechanism over trivial backgrounds. 
However, the dynamics of the scalar field is completely different 
from the usual approach. The action for the scalar field has a pure 
topological contribution from the background. The mass term becomes 
increased because of this contribution. There is also a constraint on 
the scalar field which severely restricts its dynamics. When the 
background is switched to zero the standard formulation is regained. 

Finally, the approach shows how topological sectors, in these cases 
related to $BF$ topological theories, may be incorporated to the 
standard Higgs mechanism. It is expected that the same approach  will 
lead to the non-Abelian massive gauge theories in three dimensions. 
This is a very interesting case since there are two massive 
non-equivalent gauge invariant actions in three 
dimensions{\cite{re:(7) Pio-Leal-alvaro}}, probably corresponding to 
the weak and strong coupling regimes of the same theory. It would be 
interesting to see if the Higgs mechanism relates in any way both 
models. Another interesting point  would be to analyze if  the same 
approach may introduce topological sectors to the low energy 
effective actions of supersymmetric theories.

\vskip .5cm
\noindent{\Large{\bf Acknowledgement}}
\vskip .5cm

We would like to thank P. Arias and M. I. Caicedo for interesting 
discussions. This research is supported by {\it Decanato de 
Investigaciones de la Universidad sim\'on Bol\'{\i}var} Proyecto 
USB-DID/G11.

\end{document}